\documentclass{article}
\usepackage{spconf,amsmath,graphicx}
\usepackage{multirow}
\usepackage{tabularx,lipsum,xcolor}
\usepackage{url,cite}
\usepackage[breaklinks,colorlinks]{hyperref}
\newcolumntype{Y}{>{\centering\arraybackslash}X}
\usepackage[subtle]{savetrees}
\usepackage{adjustbox}

\newcommand{\newpara}[1]{\vspace{4pt}\noindent\textbf{#1}}

\title{Rethinking Session Variability: Leveraging Session Embeddings for Session Robustness in Speaker Verification}
\name{ \begin{tabular}{c} Hee-Soo Heo$^{1}$,  KiHyun Nam$^{2}$, Bong-Jin Lee$^{1}$, Youngki Kwon$^{1}$, \\ Minjae Lee$^{1}$,  You Jin Kim$^{1}$, Joon Son Chung$^{2}$ \end{tabular}}
\address{
  $^1$NAVER Cloud Corporation, South Korea \\
  $^2$Korea Advanced Institute of Science and Technology, South Korea}
\begin{document}
%
\maketitle
\begin{abstract}
In the field of speaker verification, session or channel variability poses a significant challenge. 
While many contemporary methods aim to disentangle session information from speaker embeddings, we introduce a novel approach using an additional embedding to represent the session information. 
This is achieved by training an auxiliary network appended to the speaker embedding extractor which remains fixed in this training process.
This results in two similarity scores: one for the speakers information and one for the session information.
The latter score acts as a compensator for the former that might be skewed due to session variations.
Our extensive experiments demonstrate that session information can be effectively compensated without retraining of the embedding extractor.

\end{abstract}
\section{Introduction}
\label{sec:intro}

In the evolving domain of speech processing, speaker verification plays a crucial role, having various real-world applications ranging from voice-based security systems to personalised speech assistants. 
Central to robust speaker verification is the extraction of speaker embeddings, which encapsulate the unique characteristics of an individual's voice~\cite{dehak2010front, variani2014deep, jung2019rawnet}.
However, these embeddings are susceptible to extraneous information, largely influenced from the recording environment. 
Variabilities in recording devices, ambient noise, room acoustics, and other session-related factors can significantly affect the accuracy of these embeddings, creating misleading similarities even among distinct speakers in similar recording situations~\cite{lin2023robust, kang2020disentangled}.

Historically, when the i-vector approach was prevalent in the speaker embedding space, techniques such as linear discriminant analysis (LDA) and within-class covariance normalization (WCCN) were employed as countermeasures to diminish these unexpected session similarities~\cite{dehak2010front, matvejka2011full}. 
With the advances of deep learning and its application to this domain, efforts have shifted towards disentangling speaker information from session information directly within the embedding~\cite{kang2020disentangled, michelsanti2017conditional, huh2020augmentation}.
Various strategies have been studied in this direction -- while some leverage the adversarial approach, others design novel loss functions to achieve the same goal~\cite{mohammadamini2022learning}. 
However, a clear problem with these methods is that while trying to separate session-related information from speaker-specific details, important characteristics of the speaker might be lost. 
In simpler terms, in the process of removing unwanted session information, one might also unintentionally remove features that help identify the speaker.

In light of these challenges, this paper introduces an alternative approach. 
Instead of disentangling session-related information from the embedding, we present a framework to compensate for it at the score level. 
Our methodology capitalises on the use of an auxiliary network, seamlessly appended to the original speaker embedding extractor. 
The auxiliary network is designed to represent session information found within speaker embeddings. 
A key facet of our framework ensures that the primary speaker embedding extractor remains fixed during this process.
Consequently, our system yields a twofold output; a similarity score reflecting speaker characteristics and another gauging session attributes. 
The latter, acting as a compensator, has the potential to rectify any discrepancies in the speaker score induced by analogous or differing session conditions. 
Our empirical evaluations, spanning various model architectures and evaluation configurations, underscore the feasibility of session compensation without the need for retraining the original embedding extractor.

The paper is organised as follows. 
Section~\ref{sec:framework} introduces the proposed framework. 
Experiments and result analysis are presented in Section~\ref{sec:exp}, followed by conclusion in Section~\ref{sec:con}.

\begin{figure*}[t!]
    \centering
    \includegraphics[width=\linewidth]{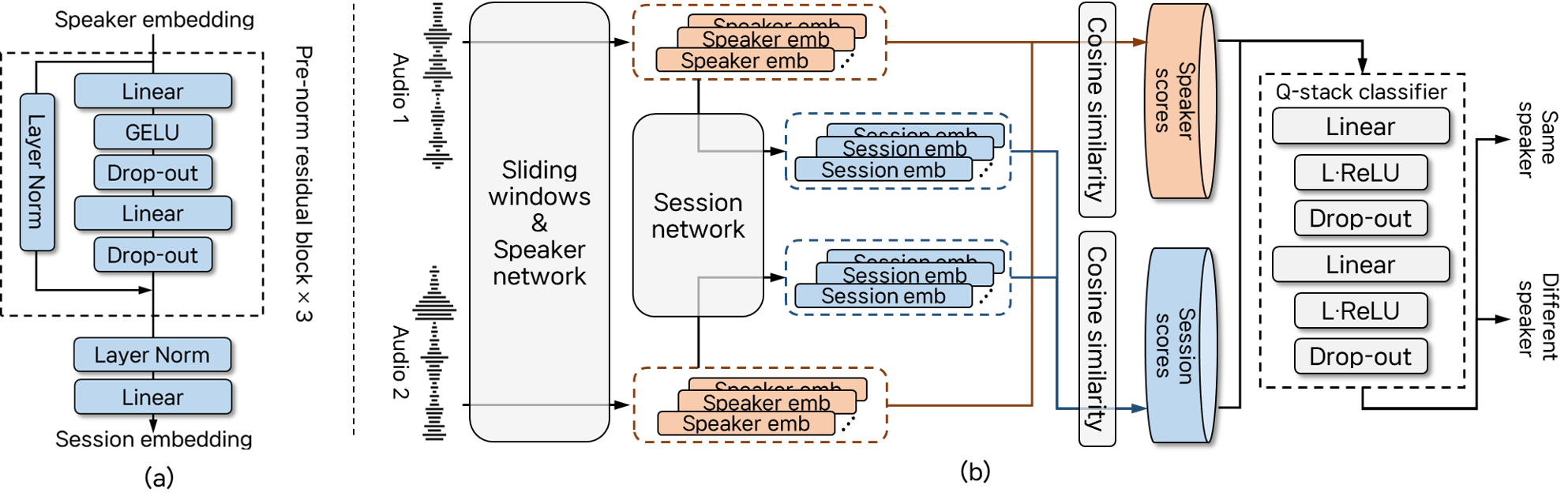}
    \caption{The session compensating framework for speaker verification. (a) Illustration of the session network. The network receives speaker embeddings and, via multiple pre-norm residual blocks, produces session embeddings. (b) Outline of the speaker verification process within the proposed framework: Upon receiving two utterances, the system extracts corresponding session and speaker embeddings. Similarities between these embeddings are then calculated. The computed similarities are subsequently input into the Q-stack classifier to determine whether the two utterances originate from a same speaker or two distinct speakers.}
    \label{fig:framework}
    \vspace{-5pt}
\end{figure*}

\section{Framework for Session Variability Compensation}
\label{sec:framework}
In this section, we present a novel framework specifically designed to address and compensate for session variability in speaker verification tasks. 

\subsection{Speaker Embedding Extraction}

For this study, we leverage pre-trained embedding extractors, drawing from methods that have proven efficacy in conventional recipes.
Specifically, we have evaluated three models that represent a diverse cross-section of state-of-the-art architectures. 
These models are ECAPA-TDNN~\cite{desplanques2020ecapa}, RawNet3~\cite{jung2022pushing}, and MFA-Conformer-based speaker embedding extractors~\cite{zhang2022mfa, jung2022large}.

\subsection{Session Embedding Extraction}

Within the domain of speaker verification, speaker embeddings efficiently capture the intrinsic attributes of a speaker's speech.
However, these embeddings may also contain subtle information specific to the recording session, like background noise or recording device characteristics. 
Recognising the need to isolate such session-specific nuances from the core speaker features, we introduce the session network.

\newpara{Network architecture.} 
This network is attached to the speaker embedding network. 
Simplistically composed of several fully-connected layers, drop-out and GELU activation~\cite{hendrycks2016gaussian}, the session network's primary role is to extract session information contained within the speaker embedding. 
Figure~\ref{fig:framework}-(a) shows the detailed composition of the session network. 
It's designed to differentiate between the inherent speaker characteristics and the variabilities introduced by different recording sessions.

\newpara{Training strategy.} 
For effective extraction of session information, it's paramount to train the network using a specially designed loss function. 
In addition, utilising datasets such as VoxCelebs~\cite{Nagrani19,chung2018voxceleb2}, which offers multiple sessions for individual speakers, is essential. 
For the session network, the training data comprises pairs – both positive and negative – drawn from the VoxCeleb datasets.
These pairs are constructed by pairing two utterances. 
First, utterances for a positive pair stem from a same session and a same speaker, with identical augmentation techniques applied. 
This setup ensures that any discrepancy in the embeddings is predominantly due to session variations.
Conversely, a negative pair includes two utterances from the same speaker but from different sessions, with distinct augmentations applied. 
This highlights the impact of session differences manifested within speaker embeddings.
To elaborate further, consider a speaker denoted as $i$, randomly selected from our dataset. 
For our training, we aim to consider the speaker's utterances across two distinct sessions. 
Thus, for each chosen session, two random utterances are selected. 
This process gives us a notation, $u_{i,s,u}|s \in \{0,1\}, u \in \{0,1\}$, where $i$ stands for the selected speaker, $s$ denotes the session and $u$ indicates the utterance.
Now, for a definition of the loss function, we consider all possible combinations of sessions ($s$) and utterances ($u$). 
Our objective is to compute a loss value, $\mathcal{L}$, which would measure the difference or similarity between these combinations. 
This loss is determined as:
\begin{equation}
\mathcal{L}=
\begin{cases}
        1 - S(se(u_{i,s1,u1}), se(u_{i,s2,u2})), & \text{if } s1 == s2 \\
        S(se(u_{i,s1,u1}), se(u_{i,s2,u2})), & \text{otherwise }
\end{cases}
\end{equation}
\noindent where $S(\cdot,\cdot)$ is a function indicating cosine similarity between two embeddings and $se(u)$ is session embedding from utterance $u$.
It's worth noting that we do not consider pairs from different speakers while training the session network, ensuring the focus remains strictly on session variability. 
The session information is directly inferred from the video ID in the VoxCeleb datasets. 
In our context, two utterances are considered to be from the same session if they originate from an identical video.

\subsection{Speaker Verification Using the Proposed Framework}

In this section, we present our speaker verification procedure underpinned by our novel framework.
In our study, we consider each verification trial to be constructed from a pair of utterances. 
From each of these utterances, two types of embeddings can be extracted: one that represents the characteristics of the speaker (the speaker embedding) and another that embodies the particularities of the recording session (the session embedding).

\begin{figure}[t!]
    \centering
    \includegraphics[width=\columnwidth]{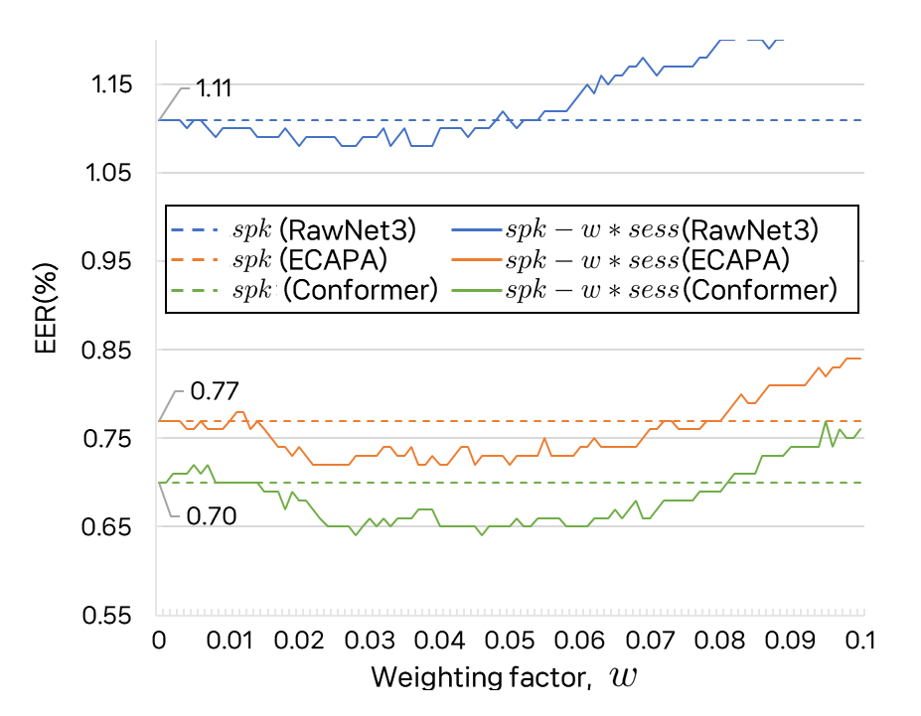}
    \caption{Variation in speaker verification performance on the original VoxCeleb1 test set for three distinct embedding extractors. The graph shows the influence of session similarity($sess$)'s weight $w$ on each extractor's performance. A clear trend emerges, highlighting the role of session similarity as a compensatory across all models evaluated.}
    \label{fig:exp}
    \vspace{-5pt}
\end{figure}

\newpara{Score-level compensator.} 
Once we have these embeddings, we can measure how similar they are. 
We compare the speaker embeddings from both utterances to get a ``speaker similarity'' score. 
This value essentially offers a metric that quantifies how alike the two utterances are, based on the characteristics of the speakers. 
On a parallel track, the session similarity is determined through the cosine similarity of the two session embeddings.
This similarity shows how alike the two recordings are, based just on details from the recording session.
Having obtained these similarities, the final step is to integrate them into a composite score that would be instrumental for verification. 
The formula we propose for this is:
\begin{equation}
score = spk - w * sess,
\end{equation}
where $spk$ and $sess$ indicate speaker and session similarities, respectively, and $w$ stands as a weighting factor for the session similarity.
By subtracting a weighted session similarity from the speaker similarity, we aim to rectify any biases present in the speaker similarity attributed to session-related variations. 
Thus, the goal is to compensate for the session-induced biases, ensuring that the speaker's inherent characteristics shine through without the unexpected influence of session-specific attributes.
To discern the impact of the session similarity on speaker verification, we carried out simple experiments utilising embeddings derived from the three embedding extractors. 
The focal point of this experiment was to adjust a weight value, and subsequently, observe how it influenced the performance of speaker verification. 
We conducted our tests using the VoxCeleb1 original test set, and the results are shown in Figure~\ref{fig:exp}. 
The results reveal that simple action of subtracting the session similarity can reduce the error in speaker verification. 

\newpara{Q-stack-based compensator.} 
Nonetheless, there exists a limitation to the above approach. 
The foundational premise of the approach is predicated on the assumption that the correlation between the speaker and session similarities is linear. 
However, in practical scenarios, this relationship might exhibit a more complex nature, suggesting the necessity for a sophisticated approach to accurately compensate for these interactions.
To address this, we utilised an additional classifier which takes in both the speaker and session similarities and makes a binary decision. 
Essentially, it determines whether the two utterances originate from the same speaker or not. 
This new approach allows us to capture the non-linear relationship between the two similarities.
The concept of this classifier is derived from a framework termed ``Q-stack"~\cite{kryszczuk2007improving}. 
The Q-stack classifier is employed to process two separate sets of similarities derived from two utterances, with the primary objective of deciding whether these utterances are from an identical speaker or not.
The operation of the Q-stack-based framework is as follows.
First, it takes in $200$ similarities; half represents speaker similarities, and the other half stands for session similarities. 
These specific quantities originate from the well-known VoxCeleb trainer's recipe~\footnote{\url{https://github.com/clovaai/voxceleb_trainer}}.
This procedure extracts $10$ embeddings from an individual utterance through a sliding window technique. 
Consequently, when comparing a pair of utterances, the possible combination results in $10 \times 10$ similarities, leading to a combined total of $100$ similarities for each type of embedding.
For a more detailed architecture of the Q-stack, it is structured with three fully-connected layers, drop-out, and non-linear activation. 
These layers consist of $400$ nodes, except the output layer with only two nodes. 
All hidden nodes are activated by leaky ReLU function for non-linearity.
Figure~\ref{fig:framework}-(b) shows the overall operation of the proposed framework, including the structure of the Q-stack classifier.

\begin{table*}[t]
\centering
\caption{A comparison of the performances using different models and evaluation sets. 
``Baseline" shows results from the usual speaker embedding. 
``Score comp" shows the outcomes when session variability is compensated at the score level. 
``Q-stack" denotes results when session variability is addressed using session embedding complemented by an additional classifier.} 
\label{tab:main}
\begin{adjustbox}{max width=\textwidth}
\begin{tabular}{l|cccc|cccc|cccc}
\hline
\hline
\multicolumn{1}{c}{\multirow{2}{*}{EER(\%)}} & \multicolumn{4}{|c|}{RawNet3}                                                                                 & \multicolumn{4}{c|}{ECAPA-TDNN}                                                                                   & \multicolumn{4}{c}{Conformer}                                                                               \\
\multicolumn{1}{c}{}                         & \multicolumn{1}{|l}{Vox1-O} & \multicolumn{1}{l}{N-SRE} & \multicolumn{1}{l}{VN-Mix} & \multicolumn{1}{l|}{VC-Mix} & \multicolumn{1}{l}{Vox1-O} & \multicolumn{1}{l}{N-SRE} & \multicolumn{1}{l}{VN-Mix} & \multicolumn{1}{l|}{VC-Mix} & \multicolumn{1}{l}{Vox1-O} & \multicolumn{1}{l}{N-SRE} & \multicolumn{1}{l}{VN-Mix} & \multicolumn{1}{l}{VC-Mix} \\
\hline
Baseline                                     & 1.11                       & 13.52                    & 10.51                    & 3.32                      & 0.77                       & 11.29                    & 6.90                     & 2.17                      & 0.70                       & 8.70                     & 3.48                     & 1.99                     \\
\hline
Score comp                                   & 1.12                       & 13.33                    & 8.91                     & 3.05                      & 0.75                       & 10.92                    & 5.84                     & 2.02                      & 0.69                       & 8.58                     & 3.43                     & 1.88                     \\
Q-stack                                      & 1.06                       & 12.98                    & 7.34                     & 3.03                      & 0.71                       & 10.64                    & 4.22                     & 1.98                      & 0.65                       & 8.39                     & 3.34                     & 1.51                    \\
\hline
\end{tabular}
\end{adjustbox}
\end{table*}

\begin{table}[t]
\centering
\caption{A comparison of the effect of the ensemble methods. ``Single best" shows the top-performing model on its own. ``Averaging scores" displays results when we combine scores from several models the usual way. ``Proposed" gives results using our new ensemble method with Q-stack.}
\label{tab:ensemble}
\begin{tabular}{lcccc}
\hline
\hline
EER(\%)          & \multicolumn{1}{l}{Vox1-O} & \multicolumn{1}{l}{N-SRE} & \multicolumn{1}{l}{VN-Mix} & \multicolumn{1}{l}{VC-Mix} \\
\hline
Single best      & 0.70                       & 8.70                     & 3.48   & 1.99                     \\
Averaging scores & 0.63                       & 8.88                     & 5.16   & 1.97                        \\
Proposed          & 0.56                       & 8.14                     & 3.17  & 1.44                       \\
\hline
\end{tabular}
\end{table}

\section{Experiements}
Experiments were conducted to evaluate the proposed speaker verification framework on four independent datasets.
The first two subsections describe implementation details and evaluation protocols across all experiments, while subsequent subsections describe experiments on the proposed system configuration.
\label{sec:exp}

\subsection{Implementation details}
For the evaluation of the proposed system, various datasets and models were employed. 
We selected multiple datasets for the training process: VoxCeleb1\&2~\cite{Nagrani17,chung2018voxceleb2}, VOiCES~\cite{richey2018voices}, CommonVoice~\cite{ardila2020common} and telephone speeches from NIST SRE corpora. 
ECAPA-TDNN and RawNet3 models were trained using the VoxCeleb1\&2 datasets.
The Conformer-based system was trained leveraging the VoxCeleb1\&2, NIST SRE 2004, 2006, and 2008~\cite{przybocki2007nist, martin2009nist}, and CommonVoice datasets.
The Q-stack system, distinctively, was trained on the test set of the VOiCES dataset.
For augmentation, we use reverberations and noises from simulated RIRs and MUSAN datasets~\cite{Ko2017ARecognition,Snyder2015MUSAN:Corpus}. 
Augmentation configurations follow that of \cite{heo2020clova}.

\subsection{Evaluation protocol}
We evaluated performance using the VoxCeleb1 original test set (Vox1-O), 10sec-10sec
protocol of NIST SRE 2010 evaluation (N-SRE)~\cite{martin2010nist}, and unique combined datasets. 
The initial evaluation of our system was carried out using two primary datasets: Vox1-O and N-SRE. 
These datasets contain audio data from varied sources and were chosen because they internally include session variability. 
To further evaluation, we introduced two custom datasets, VN-Mix and VC-Mix, crafted to test the systems' performance under challenging scenarios.
First, VN-Mix (VoxCeleb and NIST) was composed of trials from Vox1-O and N-SRE. 
A notable aspect of this combination is the intrinsic domain difference between the two datasets.
Specifically, N-SRE includes telephone speech while Vox1-O contains YouTube video clips. 
Given this contrast in source domains, it's hypothesised that a similarity bias might arise due to these inherent differences. 
For VC-Mix (VoxCeleb and VoxConverse), we combined positive pairs from Vox1-O with negative pairs from the ``single" protocol of VoxConverse~\cite{chung2020spot}, as referenced in~\cite{jung2023search}.
The positive pairs from Vox1-O comprise utterances from multiple sessions. 
In contrast, the negative pairs from VoxConverse are restricted to a singular session. 
This composition suggests the challenge, presenting both hard positive and negative pairs. 
In simple words, VC-Mix combines two types of pairs: one with the same speaker from different sessions and another with different speakers from a single session.
The structure of VC-Mix was inspired by the dataset used in the VoxSRC 2022 challenge~\cite{huh2023voxsrc}.
All telephone speech is up-sampled from 8kHz to 16kHz. 
The performance metric used to compare the models' performance was the well-known equal error rate (EER).

\subsection{Comparison with single system}
In Table \ref{tab:main}, we presented a comprehensive comparison of the baseline system against our proposed systems across varied models and evaluation datasets. 
A key observation was the robustness and enhancement in EER offered by the proposed systems, which use session embeddings. 
Focusing on the ``score comp" row, the results show the positive impact of session compensation using equation (2).
The value of the weighting factor $w$ was determined using test trials from the VOiCES dataset.
Furthermore, the ``Q-stack" row introduces further improvement from an additional classifier. 
This suggests that the classifier helps model a non-linear relationship between session and speaker similarities.

\subsection{Comparison with ensemble system}
Table \ref{tab:ensemble} shows the impact of different ensemble techniques on model performance. 
A conventional ensemble averages multiple scores from various models. 
However, with our Q-stack system, this ensemble is more sophisticated. 
Instead of merely averaging, it inputs scores from different models in unison.
In particular, we increased the number of input scores from $200$ to $600$ when combining the three models.
The experimental results highlighted the superior performance of the Q-stack-based ensemble, especially on the N-SRE dataset and the VN-Mix containing the corresponding dataset. 
Conventional ensemble techniques, on the other hand, exhibited a decrement in performance on the N-SRE dataset, attributed to some models' limited exposure to telephone speech during their training. 

\section{Conclusion}
\label{sec:con}

In the domain of speaker verification, session variability is a well-known factor that can lead to performance degradation. 
Traditional methods often aim to modify or enhance the speaker embedding to handle this issue. 
Contrary to this, we suggest a novel approach; rather than adjusting the speaker embedding, we propose that session information should be treated as a separate entity.
Comprehensive experiments, spanning a variety of models and datasets, demonstrate that the proposed method not only mitigates the effects of session variability but also has valuable implications for model ensemble and score calibration.

\clearpage
\ninept
\bibliographystyle{IEEEbib}
\bibliography{shortstrings,mybib}

\end{document}